\documentclass[twocolumn,showpacs,preprintnumbers,amsmath,amssymb,prb]{revtex4}
\usepackage{graphicx}
\usepackage{dcolumn}
\usepackage{bm}

\begin{document}

\title{Energy levels and decoherence properties of  single electron and nuclear\\ spins in a defect center in diamond}
\author{I. Popa, T. Gaebel, M. Domhan, C. Wittmann, F. Jelezko, and J. Wrachtrup}

\affiliation{3. Physikalisches Institut Universit\"{a}t Stuttgart}
\date{\today}

\begin{abstract}
The coherent behavior of the single electron and single nuclear spins of a defect center in diamond and a $^{13}C$ nucleus in its vicinity, respectively, are investigated. The energy levels associated with the hyperfine coupling of the electron spin of the defect center to the $^{13}C$ nuclear spin are analyzed. Methods of magnetic resonance together with optical readout of single defect centers have been applied in order to observe the coherent dynamics of the electron and nuclear spins. Long coherence times, in the order of $\mu$s for electron spins and tens of $\mu$s  for nuclear spins, recommend the studied system as a good experimental approach for implementing a 2-qubit gate.
\end{abstract}

\maketitle

Long decoherence time and a precise understanding and manipulation of state evolution are one of the most important requirements to be met by a quantum system in order to implement quantum computing algorithms \cite{vin95}. For this reasons among all the solid state systems considered for quantum computing, spins are the most promising in respect to long decoherence times and state control. An efficient system for this purpose is given by single paramagnetic defect centers in diamond \cite{gruber97}.  Recently, single spin readout in Nitrogen-Vacancy (N-V) defects in diamond was demonstrated \cite{jelezko1}. The coherent evolution of the electron spin of a defect center in diamond was previously reported \cite{jelezko2} and a two qubit conditional quantum gate was demonstrated \cite{jelezko3}. Here, we extend our investigation to the system composed of a single electron spin hyperfine coupled to two nuclei, one $^{13}C$ and a $^{14}N$. 

\begin{figure}
\includegraphics{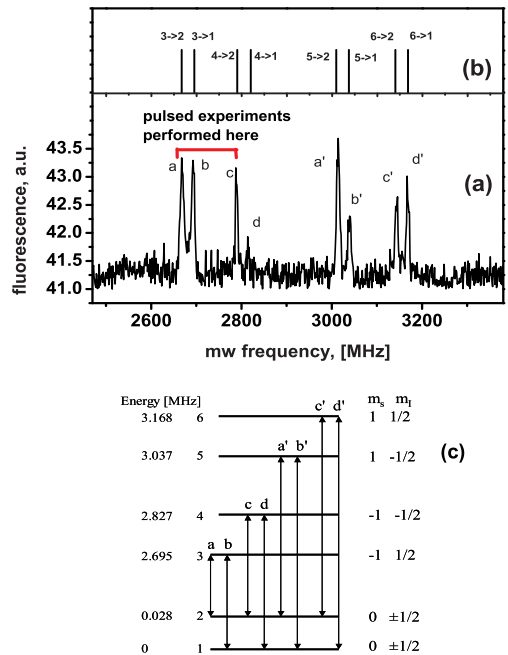}
\caption{\label{fig:epsart1} a) The experimental cw ODMR spectrum recorded for an N-V center coupled to $^{13}C$. b) The stem plot shows the calculated transitions strengths and the frequencies at which they occur; the transitions strengths are equal, without considering any effects related to optical readout. c)	Energy level scheme of the N-V center coupled to a $^{13}C$ nucleus, calculated with the Hamiltonian presented in the text; the spin functions are of the form $|m_s m_I\rangle$. Transitions are indicated with arrows and identified correspondingly in the cw spectrum in \ref{fig:epsart1}(b).}
\end{figure}

\begin{figure}
\includegraphics{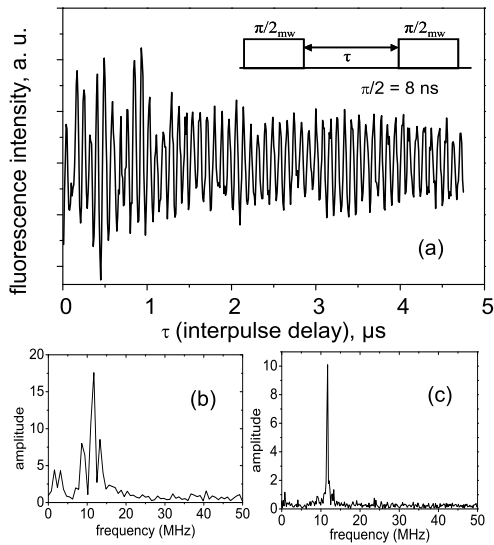}
\caption{\label{fig:epsart2} a)	The FID on the electron spin. The inset to a) shows the microwave pulse sequence employed. There is no visible decay of the oscillations in the time range considered. For short inter-pulse delays, a modulation due to $^{14}N$ can be observed. The FID measurement was recorded for a different magnetic field magnitude than that was for recording the cw spectrum in Fig.~\ref{fig:epsart2}(b)
b)	Fourier transform of the FID data; the line at 12 MHz corresponds to the splitting between the levels 1 and 2 (see Fig.~\ref{fig:epsart1}(a), while the less intense lines are due to the $^{14}N$ coupling. The Fourier transform was performed on the data for inter-pulse delays up to 1.5 $\mu$s. 
c)	Fourier transform on the data corresponding to inter-pulse delay higher than 1.5 $\mu$s. The satellite lines in the previous time range are not present anymore due to the fast decay of the $^{14}N$ induced-modulation of the FID. 
}
\end{figure}

The Nitrogen-Vacancy (N-V) defect center in diamond is a paramagnetic system $(S = 1)$ consisting of a substitutional $^{14}N$ atom next to a vacancy into an adjacent lattice site. The N-V defect center occurs naturally in diamonds containing substitutional nitrogen. The ground and first excited states ($^3A$ and $^3E$, respectively) of the defect are spin triplets. The center exhibits a strong dipole allowed optical transition between the $^3E$ and $^3A$ states, at 637 nm \cite{jelezko2}, allowing for its optical detection as a single center.  In the absence of an external magnetic field, the ground state is split in the crystal field into a singlet state Z $(m_s = 0)$ and a doublet X, Y $(m_s = \pm1)$, separated by 2.88 GHz. Optically Detected Magnetic Resonance (ODMR) is performed in the continuous wave regime by monitoring the changes in the fluorescence intensity emitted by the optically excited N-V center, upon sweeping the microwaves. At the mw resonance value, 2.88 GHz in zero field, the populations of the ground state spin sublevels will change, leading to a drop in the fluorescence signal (negative ODMR effect) \cite{niz}.   
Experimental work was done with a home-built setup. The confocal microscope operates over a wide range of temperatures (2 to 300 K). Microwaves are transmitted to the sample using an ESR microresonator (provided by D. Suter of University Dortmund).  The small size of the diameter of the loop (500 $\mu$m) allows to achieve high values for the microwave Rabi frequencies, up to 50 MHz . The sample material consists of nanocrystals  made out of  type 1b diamond . The subwavelength size of the diamond nanocrystals prevents eventual losses of fluorescence via internal reflection on the surface that occur in the case of using a large diamond. 
Hyperfine coupling of the electron spin occurs to $^{13}C$ in the surrounding lattice and to $^{14}N$ at the defect. For nearest neighbor carbons the hyperfine coupling is around 130 MHz, for second shell neighbors it is 70MHz and less than 10 MHz for the third shell \cite{abinitio,loubser}. 
The general Hamiltonian for the electron spin coupled to a nucleus is given by: 
\begin{eqnarray}
{\cal H}= D\left(S_z^2-\frac{1}{3}S^2\right)+\beta_e\bm{B}\stackrel{=}{g_e}\bm{S}+\bm{S}\stackrel{=}{A}\bm{I}\nonumber\\
+P\left[I_z^2-\frac{1}{3}I^2\right]\nonumber
\end{eqnarray}

,where D = 2.88 GHz is the zero-field splitting, $\bm{B}$ is the applied external magnetic field, $\stackrel{=}{A}$  is the hyperfine coupling tensor between the electron spin and the nuclear spin, and P is the quadrupole contribution for nuclei with $I\geq1$ . For large magnetic fields $(B>1 T)$ the electron Zeeman term is by far the most important contribution in the spin Hamiltonian. The spin is then quantised along the direction of the external B field. In this case a first order perturbation approach is sufficient to explain the energy level diagram of the electron spin. However, in the present experiments values for B between 0.008 T and 0.01 T were chosen. Partly this was because our present set up does not allow for larger fields but it also turned out that the ODMR effect, i.e., the decrease in fluorescence due to microwave resonance, can decrease at higher B fields. For such low values of B the Hamiltonian needs to be diagonalized numerically in order to extract the transition frequencies. Fig.~\ref{fig:epsart1} compares the simulated stick spectrum (Fig.~\ref{fig:epsart1}(b)) with an experimental spectrum of a defect center where hyperfine coupling of the electron spin with a single $^{13}C$ nuclear spin in the first coordination shell is measured. The orientation and amplitude of the external magnetic field were used as fit parameters in the calculations, as the actual state of the experimental setup does not allow for their direct determination. The magnitude of the magnetic field for the spectrum shown in Fig.~\ref{fig:epsart1}(a) was 140 Gauss, oriented at an angle of 26$^o$ relative to the $C_3$ symmetry axis of the defect center. With these fit parameters the calculated spectrum reproduces the measured one quite well. Fig.~\ref{fig:epsart1}(c) shows the calculated energy level scheme of the system composed of the electron spin of the N-V hyperfine coupled to the $^{13}C$ nucleus. Levels 1 and 2, the lowest energy levels, are linear combinations of $\left|0~-1/2\right\rangle$ and $\left|0~1/2\right\rangle$, where the eigenfunctions are in the form $\left|m_s m_I\right\rangle$. Theoretical calculations show that coefficients of both spin states are correspondingly equal in levels 1 and 2. This will lead to equal transition probabilities, as shown in the stick plot in Fig.~\ref{fig:epsart1}(b).  However, as can be observed from the cw spectrum in Fig.~\ref{fig:epsart1}(a), the transition lines do not have equal amplitudes. One possible reason for this might be that in the evaluation of the transition strengths, the optical readout scheme and its effects were not accounted for. The separation between states 1 and 2 is 28 MHz in Fig. 1b and corresponds to the so-called pseudo-nuclear Zeeman effect \cite{hyperfine}. For a field strength of 140 G one would expect a splitting of 0.7 MHz between level 1 and 2 by a pure nuclear Zeeman effect. The much larger splitting is due to cross terms between the electronic Zeeman interaction and the magnetic hyperfine interaction. The splitting between levels 3 and 4 correspond entirely to hyperfine coupling, since for the given amplitude of the external magnetic field, the nuclear Zeeman effect is much smaller than the hyperfine coupling. Each of levels 3 and 4 can be described by a pure high-field nuclear spin function, corresponding to two nuclear spin projections. This will be of importance for the interpretation of the electron-nuclear double resonance experiments described below.  
The coherent behavior of the electron spin was first probed by performing a free induction decay (FID) measurement. The applied pulse sequence was ${\frac{\pi}{2}}_{mw}-\tau-{\frac{\pi}{2}}_{mw}$, with $\tau$ variable. Due to the optical pumping, the system is mostly polarized in the level 1. The first mw pulse, used to convert populations into coherences, was applied between levels 1 and 3. Since the readout is optical, the second mw pulse is needed for converting the coherences back into populations. The duration of the ${\frac{\pi}{2}}_{mw}$ pulse was 8 ns. Fig.~\ref{fig:epsart2}(a) shows the experimental result of FID measurement. The applied external magnetic field was smaller in this case than that used for the cw spectrum (80 G). For delays $\tau$ up to 5 $\mu$s, there is no visible decay of the coherence. The recorded time was limited by the hardware configuration available. We would like to point out an important difference with FID measurements made for example in NMR. There measurements are done in the XY plane. This is why even on-resonance spins lead to a modulation pattern in the FID. In our experiments on-resonance spins are expected to result in a FID signal, which is constant as a function of $\tau$. However, also in Fig.~\ref{fig:epsart2}(a) modulations are visible.  The Rabi frequency of the applied mw pulses is significantly higher than the splitting between the states 1 and 2, thus, ESR transitions from both of these levels to level 3 are excited. The component related to the off-resonance transition, $2\rightarrow3$, will lead to the fringes in the FID pattern. In the rotating frame attached to the transition $1\rightarrow3$ , the off-resonance transition will move with an angular speed $\omega = \epsilon_2 - \epsilon_1$, where $\epsilon_2$ and $\epsilon_1$ are the energies of the levels 1 and 2, respectively. This is visible as modulations in Fig.~\ref{fig:epsart2}(a).
 Furthermore, for shorter inter-pulse delays, a fast-decaying modulation can be observed on top of the FID signal. This was attributed to the coupling of the electron spin to the $^{14}N$ nucleus. The hyperfine coupling constant corresponding to the $^{14}N$ is around 2 MHz, while the amplitude of the quadrupolar coupling is around 5 MHz. To prove the source of this modulation, a Fourier transform was performed on the FID data. Fig.~\ref{fig:epsart2}(b) shows a Fourier transform for short inter-pulse delays, where the modulation related to $^{14}N$ is present. The intense line at 12 MHz corresponds to the splitting between levels 1 and 2, while the other lines correspond to the calculated splitting induced by the additional coupling  to $^{14}N$  (4,7,14 MHz). The splitting due the $^{14}N$ was calculated using the Hamiltonian mentioned in the text, with the parameters from above.  
A Fourier transform of the data corresponding to longer inter-pulse delays (longer than 1.5$\mu$s, in Fig 2a) reveals only the single line at 12 MHz, and, as expected, no other transitions due to $^{14}N$ (Fig.~\ref{fig:epsart2}(c)).  
The most important contribution to this modulation decay is given by the quadrupole interaction. For nuclear spins $I\geq1$ , the quadrupole nuclear moment couples to the vibrations of the lattice (spin-phonon coupling) and accounts as the most important relaxation mechanism \cite{abragam}, while the magnetic spin-phonon coupling is a negligible relaxation mechanism for nuclei. 
\begin{figure}
\includegraphics{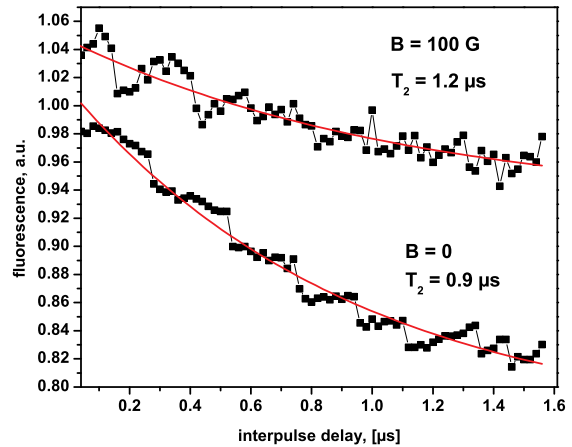}
\caption{\label{fig:epsart3}Hahn echo decay on a single electron spin for different values of the external magnetic field. The Hahn echo corresponding to zero-field decays faster due to cross-relaxation. }
\end{figure} 
To determine $T_2$ for the single electron spin, Hahn echo decay experiments have been performed. Fig.~\ref{fig:epsart3} shows the Hahn echo decay results, obtained in the absence of an external magnetic field as well as in a small external magnetic field (100 G). The shorter decoherence time in the former case can be explained by cross relaxation between the N-V spin and the spin bath, e.g., the spins of the $P_1$-centers (substitutional nitrogen impurity $S = \frac{1}{2}$) in the lattice. In the absence of a magnetic field, the single N-V centers and the N-centers are magnetically equivalent, i.e., they have similar transition frequencies, within the transition linewidth and the relaxation occurs through spin flips with the neighboring spins. However, when a magnetic field is applied, the N-V center and the spin bath will have different resonance frequencies, and the spin flips cannot occur, since as a requirement, the energy should be conserved. Thus, a longer time is needed for reaching equilibrium between the N-V spin and the spin bath, resulting in a slower decay of the Hahn echo.  
For probing the coherent manipulation of the single $^{13}C$ spin, the levels 3 and 4, corresponding to $m_I = -\frac{1}{2}$ and $m_I = \frac{1}{2}$ respectively, have been used. In order to determine the dephasing time for a single carbon nuclear spin, a modified Hahn echo sequence was applied. 
\begin{figure}
\includegraphics{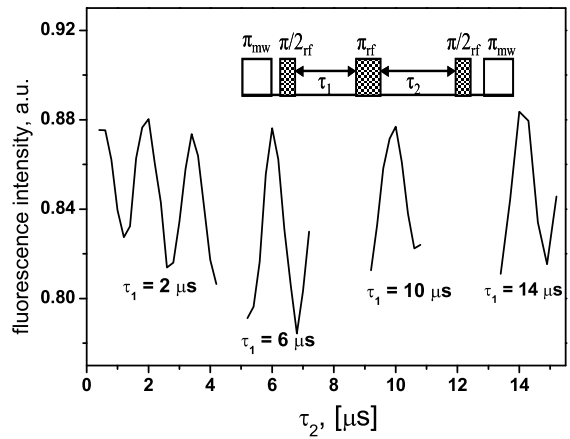}
\caption{\label{fig:epsart4} Hahn echo performed on the $^{13}C$ nucleus. The pulse sequence employed, modified accordingly for the optical readout scheme is shown in the inset to the figure. For times up to 30 $\mu$s the amplitude of the echoes show no decay.}
\end{figure} 
The coherence properties of single nuclear spin states cannot be probed directly in the actual detection configuration. Instead, the electron spin is used to intermediate the detection and manipulation of nuclear states. The applied sequence was $\pi_{mw} -{\frac{\pi}{2}}_{rf}- \tau_1 - \pi_{rf}-\tau_2-{\frac{\pi}{2}}_{rf}-\pi_{mw}$. The first mw $\pi$ pulse was used to excite the system to level 3, from where it was subjected to an rf Hahn echo pulse sequence. Fig.~\ref{fig:epsart4} shows the series of the recorded Hahn echoes for several values of the delay $\tau_1$. The amplitude of the echoes shows no decay for a time range of up to 30 $\mu$s. Nuclear spin dephasing times of up to 100 $\mu$s have been previously reported on bulk $^{13}C$ NMR measurements \cite{schaum}. Compared to the value obtained for single nuclei, it can be concluded that the hyperfine coupling to the electron spin of the N-V center does not contribute as an additional source of decoherence to the single nuclear spin. 
The N-V center in diamond provides unique opportunity to study the physics of single spins or small clusters of spins or to create certain interesting quantum states with single spins. Since we are able to precisely control the quantum state of the spins a next logical step would be to create e.g. Bells states and probe quantum correlation among the two spins. This would be to our knowledge the first test of Bells inequality with spins in solids.
\newline
\newline

This work was supported by Landesstiftung BW, and EU Project No IST-2001-37150 QIPDDF-ROSES, and DFG under the framework of project ''Quanten-Informationsverarbeitung''. One of the authors (I.P.) acknowledges the support of the Graduiertenkolleg Magnetische Resonanz Universit\"{a}t Stuttgart.

\newpage
\bibliography{paper_bib}
\end{document}